\documentclass[jap,reprint,floatfix,showpacs,showkeys,a4paper,superscriptaddress,amsmath,byrevtex]{revtex4-1}
\usepackage{graphicx,xspace,xcolor}
\usepackage{rotating}

\newcommand{\TiAlXN}[1]{Ti$_{1-x-y}$Al$_{x}$#1$_{y}$N\xspace}
\newcommand{\TizAlXN}[1]{Ti$_{z}$Al$_{x}$#1$_{y}$N\xspace}
\newcommand{\uu}[1]{\ensuremath{\,\mbox{#1}}}

\newcommand{\rev}[1]{{\color{purple}#1}}

\begin{document}
\title{Alloying-related trends from first principles: An application to the Ti--Al--X--N system}

\author{David \surname{Holec}}
\email{david.holec@unileoben.ac.at}
\affiliation{Department of Physical Metallurgy and Materials Testing, Montanuniversit\"at Leoben, A-8700 Leoben, Austria}

\author{Liangcai \surname{Zhou}}
\affiliation{Department of Physical Metallurgy and Materials Testing, Montanuniversit\"at Leoben, A-8700 Leoben, Austria}

\author{Richard \surname{Rachbauer}}
\altaffiliation{currently employed at OC Oerlikon Balzers AG, Iramali 18, LI-9496 Balzers, Liechtenstein}
\affiliation{Department of Physical Metallurgy and Materials Testing, Montanuniversit\"at Leoben, A-8700 Leoben, Austria}

\author{Paul H. \surname{Mayrhofer}}
\affiliation{Department of Physical Metallurgy and Materials Testing, Montanuniversit\"at Leoben, A-8700 Leoben, Austria}
\affiliation{Institute of Materials Science and Technology, Vienna University of Technology, A-1040 Vienna, Austria}
\affiliation{Christian Doppler Laboratory for Application Oriented Coating Development at the Department of Physical Metallurgy and Materials Testing, Montanuniversität Leoben, A-8700 Leoben, Austria}
\affiliation{Christian Doppler Laboratory for Application Oriented Coating Development at the Institute of Materials Science and Technology, Vienna University of Technology, A-1040 Vienna, Austria}

\begin{abstract}
Tailoring and improving material properties by alloying is a long-known and used concept. Recent research has demonstrated the potential of \textit{ab initio} calculations in understanding the material properties at the nanoscale. Here we present a systematic overview of alloying trends when early-transition metals (Y, Zr, Nb, Hf, Ta) are added in the Ti$_{1-x}$Al$_x$N system, routinely used as a protective hard coating. The alloy lattice parameters tend to be larger than the corresponding linearised Vegard's estimation, with the largest deviation more than $2.5\%$ obtained for Y$_{0.5}$Al$_{0.5}$N. The chemical strengthening is most pronounced for Ta and Nb, although also causing smallest elastic distortions of the lattice due to their atomic radii being comparable with Ti and Al. This is further supported by the analysis of the electronic density of states. Finally, mixing enthalpy as a measure of the driving force for decomposition into the stable constituents, is enhanced by adding Y, Zr and Nb, suggesting that the onset of spinodal decomposition will appear in these cases for lower thermal loads than for Hf and Ta alloyed Ti$_{1-x}$Al$_x$N.
\end{abstract}

\pacs{
61.66.Dk       % Alloys
68.60.Dv       % Thermal stability; thermal effects
71.15.Mb       % Density functional theory, local density approximation, gradient and other corrections
71.20.Be       % Transition metals and alloys
81.05.Je       % Ceramics and refractories (including borides, carbides, hydrides, nitrides, oxides, and silicides)
}

\keywords{Density Functional Theory; alloys; TiAlN; phase stability}
\date{\today}

\maketitle

\section{Introduction}

Ti$_{1-x}$Al$_x$N is nowadays a well-established material used as a protective hard coating due to its high hardness, relatively low coefficient of friction, and good oxidation and corrosion resistance \cite{Munz1986, PalDey2003}. As such it has attracted significant attention both in basic as well as in applied research areas, resulting also in several quantum mechanical studies on phase stability, mechanical and electronic properties of the ternary Ti$_{1-x}$Al$_x$N system \cite{Mayrhofer2006b, Alling2007a, Alling2008a, Alling2008, Tasnadi2010a, Holec2011, Holec2011a, Abrikosov2011a, ToBaben2012}.

One of the possible ways how to further tune the material properties is the concept of multicomponent alloying. For example, Y, Zr, and Hf have been shown to improve the oxidation resistance of Ti$_{1-x}$Al$_x$N \cite{Moser2007, Chen2011, Rachbauer2011b} and Cr$_{1-x}$Al$_x$N coatings \cite{Rovere2010a}, Ta to increase the high temperature durability \cite{Pfeiler2009}, Cr, Nb, and Ta to retard the decomposition process to higher thermal loads \cite{Lind2011, Mayrhofer2010, Rachbauer2011c}. The most recent successful approach is to combine theoretical studies with experimental work in order to gain deeper understanding of the experimental observations (e.g., Refs. \cite{Lind2011, Chen2011, Mayrhofer2010, Rachbauer2011b, Rachbauer2011, Rachbauer2010, Rachbauer2011c, Rovere2010a, Rovere2010, ToBaben2012}). Good examples demonstrating the ability of \textit{ab initio} calculations for guiding the experiment by predicting chemistry-related trends are e.g. Refs.~\cite{Alling2008, Rovere2010, Holec2011a} for various TM--Al--N ternary systems. However, similar systematic theoretical study showing the alloying-related trends for quaternary \TiAlXN{X} (or, more precisely, pseudo-ternary TiN--AlN--XN) systems, is missing.

In this paper we demonstrate the potential of the first principle calculations for providing such systematic and exhaustive information for Ti$_{1-x}$Al$_x$N alloyed with Y, Zr, Hf, Nb, and Ta throughout the whole quaternary phase field. It is clearly shown that with some care, the calculated results can serve as reliable ``trend-givers'' for the materials design. Lastly, it should be noted that in the paper we discuss only some related topics like the ground state properties, chemical strengthening, or onset of the spinodal decomposition; others, such as the final decomposition phase (precipitation of AlN in its stable wurtzite (B4, ZnS prototype) structure or influence on oxidation resistance, remain for the future studies.

\section{Computational approach}

We used Vienna Ab initio Simulation Package \cite{Kresse1996, Kresse1996a} employing the Density Functional Theory \cite{Hohenberg1964, Kohn1965} to perform quantum mechanical calculations. The exchange and correlation effects were described with Generalised Gradient Approximation as parametrised by Wang and Perdew \cite{Wang1991}, and implemented in projector augmented plane-wave pseudo-potentials \cite{Kresse1999}. We used the supercell approach to model random alloys, in particular $3\times3\times2$ (containing 36 atoms) supercells were constructed for the cubic rock-salt B1 structures. To do so, we optimised the short-range order parameters for pairs up to the fourth  nearest neighbour distance, for triplets up to the third nearest neighbour distance, and for quadruplets up to the second nearest neighbour distance, following the Special Quasi-random Structures methodology \cite{Wei1990}. The same technique was applied also to construct $2\times2\times2$ supercells (32 atoms) for the wurtzite B4 structures. More details on these cells can be found in Ref.~\cite{Chen2011}.

The plane wave cut-off of $500\uu{eV}$ and a minimum of $\approx3500\,\mathbf{k}\mbox{-point}\cdot\mbox{atoms}$ guarantee the total energy accuracy in the order of meV per atom. The stability of systems can be described using energy of formation, $E_f$, calculated as
\begin{multline}
   E_f=E(\mbox{\TiAlXN{X}})-\frac12\Big[(1-x-y)E(\mathrm{Ti}^{\mathrm{hcp}})\\ +xE(\mathrm{Al}^{\mathrm{fcc}})+yE(\mathrm{X}^{\xi})+\frac12E(\mathrm{N}_2)\Big]\ .
\end{multline}
Here, $E(\mbox{\TiAlXN{X}})$ is the total energy per atom of c-\TiAlXN{X}, $E(\mathrm{Ti}^{\mathrm{hcp}})$, $E(\mathrm{Al}^{\mathrm{fcc}})$ and $E(\mathrm{X}^{\xi})$ are the total energies of Ti in hexagonal close-packed (hcp, A3), Al in face-centered cubic (fcc, A1), X$=$Y, Zr, and Hf in hcp, and X$=$Nb, and Ta in body-centered cubic (bcc, A2) structures, respectively. $E(\mathrm{N}_2)$ denotes the total energy of a nitrogen molecule.

\section{Results and discussion}

\subsection{Bulk properties and phase stability of quaternary Ti--Al--X--N}\label{sec:bulk}

\begin{figure*}
  \centering
  \includegraphics[width=0.7\textwidth]{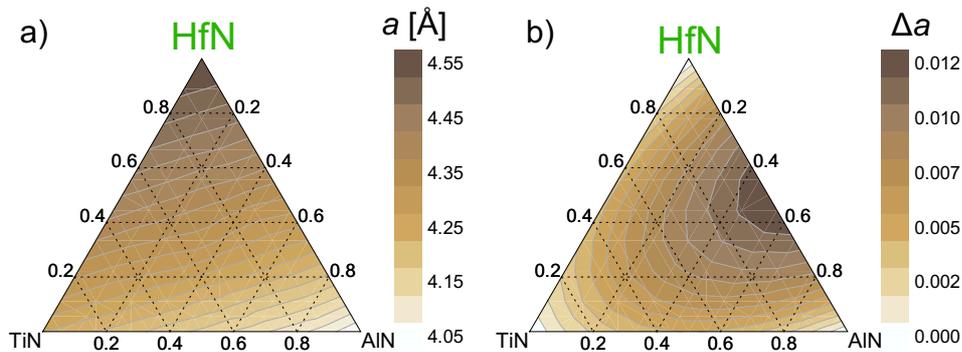}
  \caption{a) Lattice parameter, $a$, and b) its deviation, $\Delta a$, from the Vegard's estimate for the quasi-ternary TiN--AlN--HfN system.}\label{fig:aLat}
\end{figure*}

Fitting the energy--volume curve with the Birch-Murnaghan equation of state (EOS) \cite{Birch1947} yields the ground state properties: lattice parameter, $a$, total energy, $E_0$, and bulk modulus, $B$. The cubic lattice parameters of five quaternary systems investigated here as a function of AlN mole fraction, $x$, and XN mole fraction, $y$, exhibit almost linear behaviour. Straight equally spaced contours demonstrate this in Fig.~\ref{fig:aLat}a on the example of \TiAlXN{Hf}. The linearly interpolated lattice parameter according to the Vegard's empirical rule \cite{Vegard1921} reads
\begin{equation}
  a_V(x,y)=x a_{\mathrm{AlN}}+y a_{\mathrm{XN}}+ (1-x-y)a_{\mathrm{TiN}}\ .
\end{equation}
The difference $\Delta a=a-a_V$ is shown in Fig.~\ref{fig:aLat}b again for the \TiAlXN{Hf} system. It follows that the linear Vegard's estimate is erroneous by as much as $1.2\%$ in the middle of the HfN--AlN tie-line. Similarly, differences of $\approx1.5\%$, $\approx1.0\%$, and $\approx1.0\%$ are obtained also for \TiAlXN{Zr}, \TiAlXN{Nb}, and \TiAlXN{Ta} systems, respectively, with the maximum appearing always in the middle of the AlN--XN tie-line (X=Zr, Nb, Ta). \TiAlXN{Y} exhibits an error of $>2.5\%$ for Y$_{0.5}$Al$_{0.5}$N, reflecting the fact that Y is a significantly larger atom than Al. These deviations should be considered when using the Vegard's linearisation as an estimate of $a$. The reason for the non-linear behaviour of $a$ can be traced back to the gradual changes in the bonding (e.g., from ionic to metallic character), as demonstrated for Ti$_{1-x}$Al$_x$N in Ref.~\cite{Holec2011}.

In all five cases of \TiAlXN{X}, alloying an element X to Ti$_{1-x}$Al$_x$N increases the lattice constant when the Al content on the metallic sublattice, $x$, is kept constant as well as for the constant Al-to-Ti ratio, $x/(1-x-y)$. These trends are in a good agreement with previous experimental observations and theoretical studies \cite{Moser2007, Rachbauer2010, Mayrhofer2010, Rachbauer2011,Chen2011, Rachbauer2011b, Rachbauer2011c}. The calculated lattice parameters are summarised in Table~\ref{tab:data}.

\begin{figure*}
  \centering
  \includegraphics[width=\textwidth]{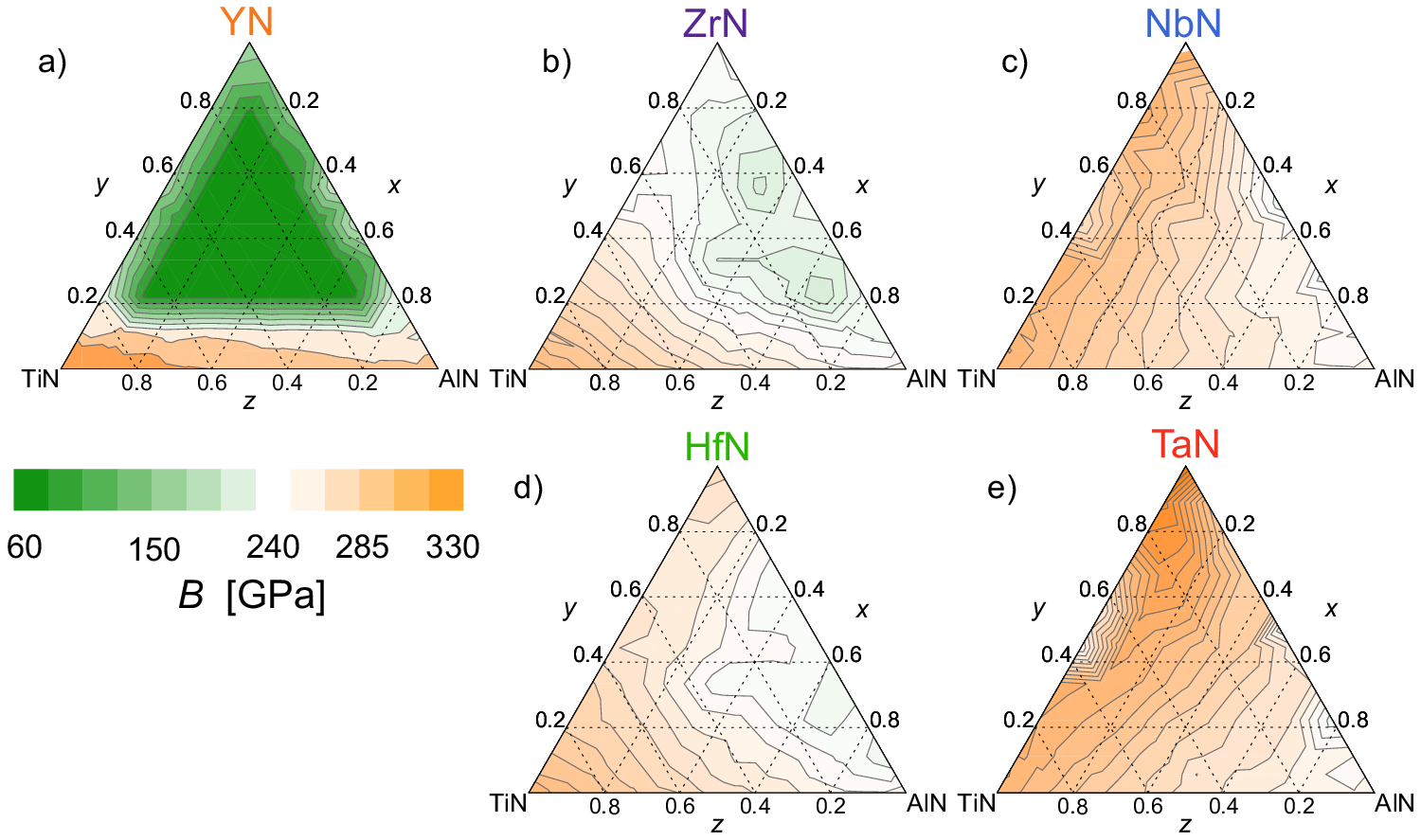}
  \caption{Bulk modulus, $B$, as a function of the composition of pseudo-ternary alloys: (a) \TizAlXN{Y}, (b) \TizAlXN{Zr}, (c) \TizAlXN{Nb}, (d) \TizAlXN{Nb}, and (e) \TizAlXN{Ta}. Countours are shown every $25\uu{GPa}$ for \TizAlXN{Y} (a), and every $5\uu{GPa}$ for the other four systems (b)--(e).}\label{fig:B}
\end{figure*}

Bulk moduli, being the measure of volume compressibility (the elastic behaviour) of the system, are plotted in Fig.~\ref{fig:B}. In all cases, the variation is non-linear but smooth. The results can be classified into three different compositional dependencies, based on the valency of the alloying element X. For the group IVB elements, Zr and Hf, it seems that the main parameter controlling $B$ is the TiN mole fraction, $z=1-x-y$ (there is only a minor variation along the tie-lines with a fixed TiN mole fraction). For example, bulk modulus of \TiAlXN{Hf} varies between $\approx240\uu{GPa}$ obtained for Al-excess Hf$_{1-x}$Al$_x$N, and $\approx290\uu{GPa}$ for TiN  \cite{Holec2012} (see Fig.~\ref{fig:B}b). For the isovalent alloy \TiAlXN{Zr}, a similar behaviour is obtained (Fig.~\ref{fig:B}d). Situation is somewhat different for the \TiAlXN{Nb} and \TiAlXN{Ta} alloys (group VB), as NbN ($B=307\uu{GPa}$) and TaN ($B=330\uu{GPa}$) \cite{Holec2012}, respectively, yield the highest bulk moduli of the quaternary systems investigated (Figs.~\ref{fig:B}c and d). For these systems, the main parameter controlling $B$ is the AlN mole fraction, $x$. Finally, yet a completely different elastic response is obtained for \TiAlXN{Y} alloy ($B_{\mathrm{YN}}=130\uu{GPa}$), where the bulk modulus reaches values well below $100\uu{GPa}$ for compositions in the centre of the compositional triangle (Fig.~\ref{fig:B}a). This is likely to be related to a strong instability of such material due to huge differences in the atomic size of individual species.

A comparison of Figs.~\ref{fig:aLat}a and \ref{fig:B}d together with the data in Table \ref{tab:data} offers the possibility to design lattice matched superlattice materials (i.e. multilayer arrangements with layer thicknesses within the order of the lattice parameter) with variations in bulk modulus. For example, TiN and Hf$_{0.3}$Al$_{0.7}$N have the same equilibrium lattice parameters of $\approx4.25\uu{\AA}$, but about $40\uu{GPa}$ difference in $B$. Another example is Ti$_{0.10}$Al$_{0.38}$Zr$_{0.52}$N and TaN, both having lattice parameter $a\approx4.42\uu{\AA}$, but the respective bulk moduli being $230\uu{GPa}$ and $330\uu{GPa}$. Systems consisting of varying softer and harder components are known to be able to e.g., hinder dislocation motion, or stop or deflect propagation of cracks \cite{Zhang2010, Li2007}. Here, an additional benefit stems from expected perfect interfaces (same crystal structures, lattice-matched).

It has been shown that for cubic structured Ti$_{1-x}$Al$_x$N thin films, the hardness reaches a maximum for the highest Al contents \rev{for which the cubic structure is still maintained} \cite{Mayrhofer2003}. The chemical compositions with hardness maximum are close to Ti$_{0.34}$Al$_{0.66}$N for physical vapor-deposited coatings. These compositions provide also the highest driving force for decomposing towards their stable  constituents, cubic TiN and wurtzite AlN, via a spinodal decomposition route. The calculated energies of formation, $E_f$, of the cubic and wurtzite phases were interpolated with cubic polynomials in $x$ and $y$, as implemented in the software package Mathematica. Equating $E_f^{\mathrm{cub}}=E_f^{\mathrm{wur}}$ yields the cross-over, an estimation of the cubic and wurtzite single phase fields. Considering only small amounts of X (i.e., Y, Zr, Nb, Hf, or Ta) up to $\approx0.1$ on the metallic sublattice, Nb and Ta do not change the maximum solubility of Al in the cubic phase considerably \cite{Mayrhofer2010, Rachbauer2010}, Zr and Hf decrease it slightly \cite{Rachbauer2011b, Chen2011, Rachbauer2011c}, while Y decreases the Al solubility limit within the cubic structure drastically from $x\approx0.7$ to $x\approx0.56$ \cite{Rachbauer2011, Moser2007}. The lowering of the solubility limit in the case of Zr and Hf is a result of widening of the dual-phase region in the Zr$_{1-x}$Al$_x$N and Hf$_{1-x}$Al$_x$N phase diagrams, as discussed in Ref.~\cite{Holec2011a}. The lowered solubility limit of Y is likely to be connected with the large differences in the atomic sizes ($r_{\mathrm{Al}}=1.432\uu{\AA}$, $r_{\mathrm{Ti}}=1.448\uu{\AA}$, and $r_{\mathrm{Y}}=1.80\uu{\AA}$) destabilising the close packed rock-salt structure while favouring the more open wurtzite structure.

\subsection{Alloying effects on Ti$_{0.5}$Al$_{0.5}$N}

In order to elucidate the role of individual elements, we use c-Ti$_{0.5}$Al$_{0.5}$N as a test composition. By replacing 1 Al or 1 Ti (out of 18 on the metallic sublattice) with 1 X atom, the alloy composition is changed to Ti$_{0.500}$Al$_{0.444}$X$_{0.056}$N (Ti-excess) or Ti$_{0.444}$Al$_{0.500}$X$_{0.056}$N (Al-excess).

\subsubsection{Chemical strength and elastic distortions}\label{sec:chemical_strength}

\begin{figure}
  \centering
  \includegraphics{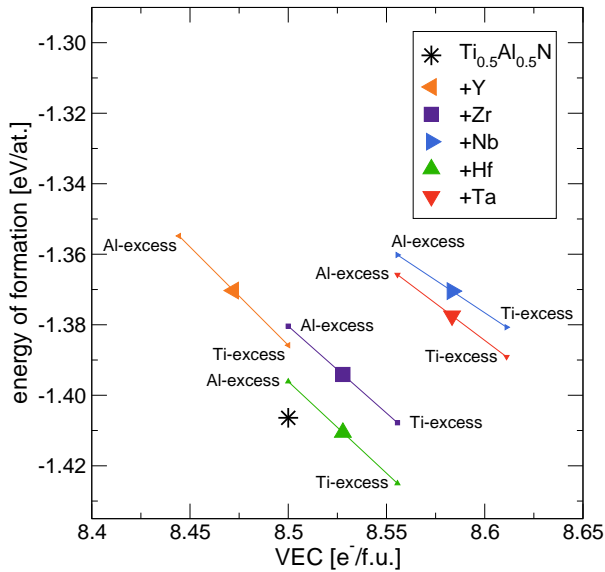}
  \caption{Energy of formation, $E_f$ as a function of the valence electron concentration (VEC) for Ti$_{0.5}$Al$_{0.5}$N, Ti-excess (Ti$_{0.5}$Al$_{0.444}$X$_{0.056}$N), and Al-excess (Ti$_{0.444}$Al$_{0.5}$X$_{0.056}$N) alloys.}\label{fig:Ef}
\end{figure}

The relative phase stability change by alloying may be quantified by energy of formation expressing the energy gain of forming an alloy of a given composition with respect to the individual elements. The results are plotted in Fig.~\ref{fig:Ef} against the valence electron concentration (VEC) which is calculated as an average value of valence electrons per formula unit (f.u., 1 metal and 1 nitrogen atom). Ti ([Ar]$3d^24s^2$), Al ([Ne]$3s^23p^1$), and N ([He]$2s^22p^3$) have 4, 3, and 5 valence electrons, respectively, hence VEC of Ti$_{0.5}$Al$_{0.5}$N is $\frac12(\frac12(4+3)+5)=8.5$. These results indicate that Y, Ta, Nb, and Zr slightly increase the energy of formation while Hf works in the opposite direction. Nevertheless, these changes are not large (below $50\uu{meV/at.}$) and could be influenced by the actual supercell configuration. In addition, one should bear in mind that these are results of $0\uu{K}$ calculations, and thus the mutual differences may change at higher temperatures due to, e.g. vibrational contribution to the Gibbs free energy.

\begin{figure}
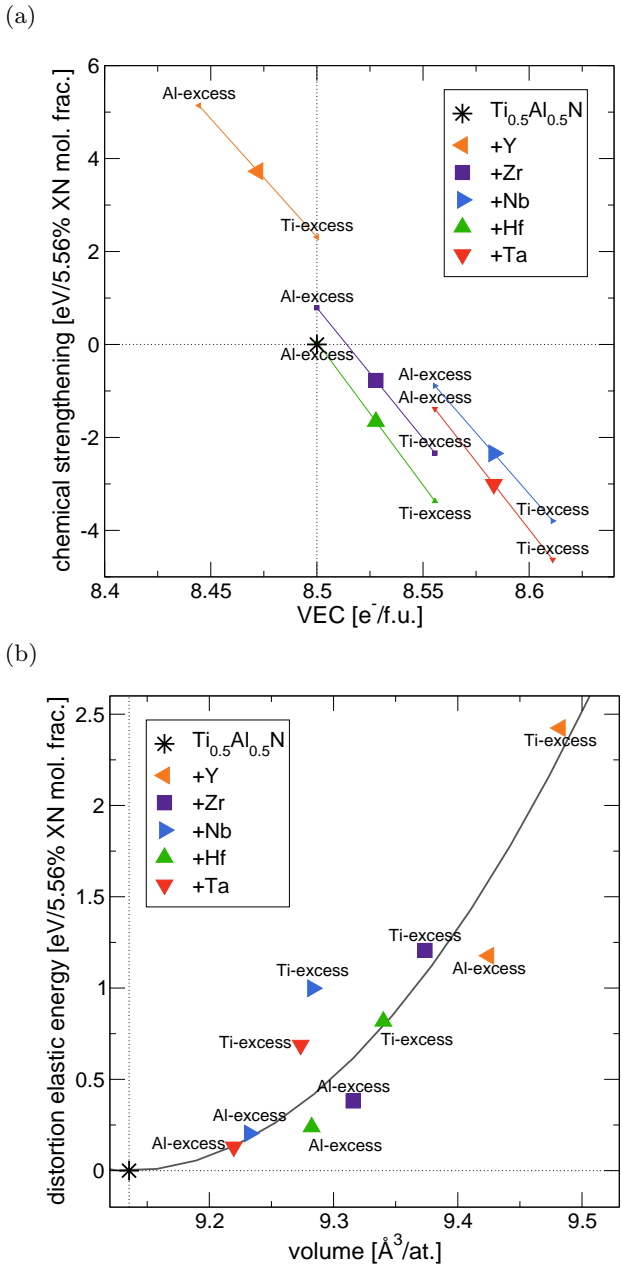

  \centering
  (a)\hfill\mbox{}\newline\newline
  \includegraphics{chemical_strength.eps}
  
  (b)\hfill\mbox{}\newline\newline
  \includegraphics{elastic_energy.eps}
  \caption{(a) Chemical strengthening and (b) alloying-related elastic energy caused by the local distortions of Ti$_{0.5}$Al$_{0.5}$N, Ti-excess (Ti$_{0.5}$Al$_{0.444}$X$_{0.056}$N), and Al-excess (Ti$_{0.444}$Al$_{0.5}$X$_{0.056}$N) alloys.}\label{fig:chem_elastic}
\end{figure}

In order to address the strengthening effect of the alloying elements, we first introduce cohesive energy calculated as 
\begin{multline}
   E_c=E(\mbox{\TiAlXN{X}})-\frac12\Big[(1-x-y)E(\mathrm{Ti})\\ +xE(\mathrm{Al})+yE(\mathrm{X})+E(\mathrm{N})\Big]\label{eq:Ecoh}
\end{multline}
where $E(X)$ is the total energy of an isolated atom $X$. Based on this quantity, two energy terms are discussed: relative chemical strength, calculated as the cohesive energy difference between relaxed Ti$_{0.5}$Al$_{0.5}$N and (unrelaxed) Ti$_{0.5}$Al$_{0.5}$N+1X, where either one Ti or Al atom (out of 18 in the supercell) is replaced with one X atom. In the second step we allow for a full structural relaxation. The total energy (or cohesive energy) decrease due to the relaxation (with fixed chemical composition of the alloy) corresponds to the elastic energy related to the local distortions caused by the foreign atom. These two contributions are shown in Figs.~\ref{fig:chem_elastic}a and b.

The relative chemical strengthening (Fig.~\ref{fig:chem_elastic}a) demonstrates the detrimental influence of Y. 
\rev{For all the other elements the relative chemical strength of (Ti$_{0.5}$Al$_{0.5})_{1-y}$X$_y$N with respect to Ti$_{0.5}$Al$_{0.5}$N becomes more negative. The cohesive energy, as defined in Eq.~\ref{eq:Ecoh}, expresses the energy (per atom) needed to de-assemble the crystalline material into individual atoms. Since the elastic relaxation of the atom positions can only lower the total or cohesive energy, a negative/positive value of the relative chemical strengthening is an indicator for stronger/weaker bonded atoms in the crystal.} This is a purely electronic effect (a local change of the electronic structure) and as such we interpreted it as \rev{electronic strengthening}. Consequently, the quaternary alloys with Zr, Hf, Nb, and Ta \rev{are expected to be more resistant against plastic deformation, leading e.g., to a change in hardness of the alloys (although there are many other aspects influencing material's hardness)}. Indeed, a comparison with the experimental values taken from the literature, $H_{\mathrm{TiAlN}}\approx33\uu{GPa}$ \cite{Chen2011}, $H_{\mathrm{TiAlYN}}\approx23\uu{GPa}$ \cite{Rachbauer2011}, $H_{\mathrm{TiAlZrN}}\approx38\uu{GPa}$ \cite{Chen2011}, $H_{\mathrm{TiAlNbN}}\approx37\uu{GPa}$ \cite{Rachbauer2011}, $H_{\mathrm{TiAlHfN}}\approx35\uu{GPa}$ \cite{Rachbauer2011b}, and $H_{\mathrm{TiAlTaN}}\approx37\uu{GPa}$ \cite{Rachbauer2011c}, supports this qualitative trend. It should be mentioned that the experimental value of $H_{\mathrm{TiAlYN}}$ corresponds to a wurtzite single phase structure, a phase change caused by the addition of $\approx10\uu{mol\%}$ YN to Ti$_{0.5}$Al$_{0.5}$N. Since the wurtzite phase is in general softer than the cubic one, this significant decrease in hardness is likely to combine both effect, the phase change as well as the weakening of the chemical bonding due to Y. Finally, the linear guide-for-the eye suggest that the chemical strength increases with increasing VEC, at least in the investigated region. 

Figure \ref{fig:chem_elastic}b shows how much elastic energy is introduced by local distortions into the material by alloying. The elastic energy is plotted against the volume per atom. The volume expansion in the first approximation equals to $3\varepsilon$, where $\varepsilon$ is the homogeneous isotropic strain. Assuming the elastic constants are not significantly influenced by alloying, the elastic energy is proportional to $\varepsilon^2$. The quadratic fit gives qualitatively a good description as can be seen in Fig.~\ref{fig:chem_elastic}b. Taking into account the \rev{metallic} radii of individual metal atoms, $r_{\mathrm{Al}}=1.432\uu{\AA}$ and $r_{\mathrm{Ti}}=1.448\uu{\AA}$, allows to explain why the smallest elastic deformations are caused by alloying with Nb ($r_{\mathrm{Nb}}=1.46\uu{\AA}$) and Ta ($r_{\mathrm{Ta}}=1.46\uu{\AA}$), followed by Hf ($r_{\mathrm{Hf}}=1.564\uu{\AA}$) and Zr ($r_{\mathrm{Zr}}=1.59\uu{\AA}$), and the largest distortions being caused by alloying with Y ($r_{\mathrm{Y}}=1.80\uu{\AA}$), as in this sequence the atomic radii increase.

In conclusion, Ta and Nb are predicted to be the most dominant strengthening elements, which at the same time introduce also the smallest lattice distortions.

\subsubsection{Density of states}

\begin{figure}
  \centering
  (a)\hfill\mbox{}\newline\newline
  \includegraphics{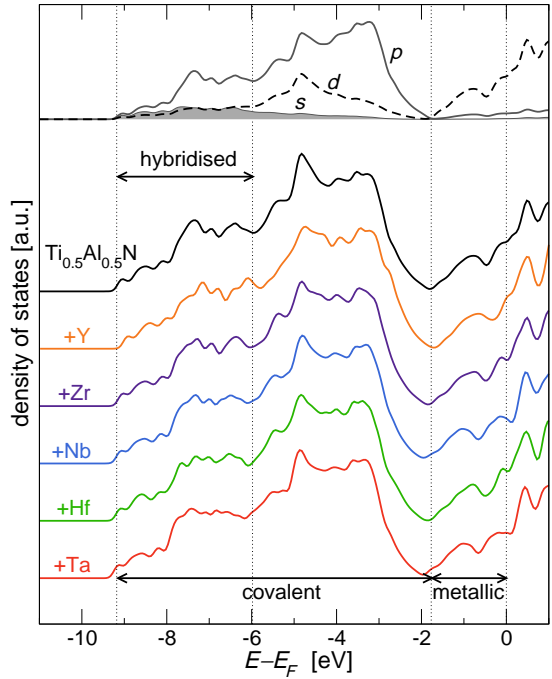}

  (b)\hfill\mbox{}\newline\newline
  \includegraphics{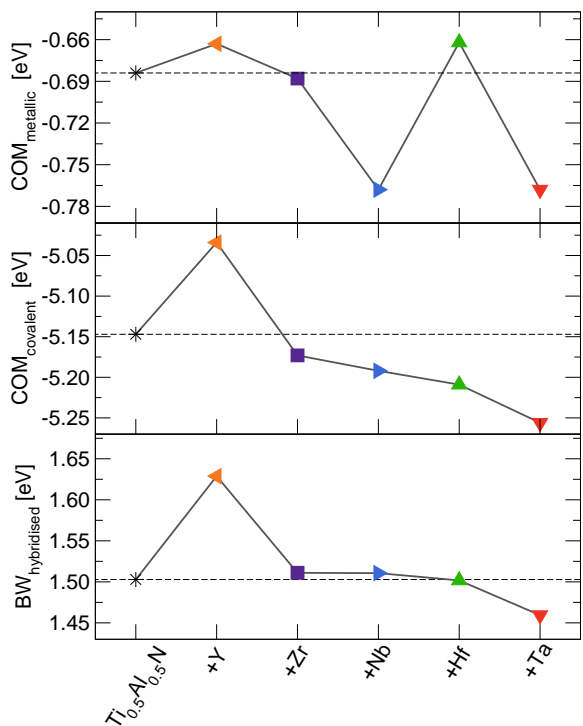}
  \caption{(a) Density of states for pure, Y-, Zr-, Hf-, Nb-, and Ta-containing Ti$_{0.5}$Al$_{0.5}$N for the Al-excess compositions ($y=0.056$). \rev{The $s$-, $p$-, and $d$-projected density of states of Ti$_{0.5}$Al$_{0.5}$N is shown on the very top.} $E_F$ denotes the Fermi energy. (b) Calculated corresponding centres of mass for metallic and convalent regions\rev{, and the band width of the $sp^3d^2$ hybridised region ($-10$ and $\approx-2\uu{eV}$)}.}\label{fig:dos}
\end{figure}

The total density of states (DOS) for Ti$_{0.5}$Al$_{0.5}$N and corresponding Al-excess alloys is shown in Fig.~\ref{fig:dos}a. The top of the valence band, as shown, can be divided into three distinct regions: \rev{(i) $-10$ to \rev{$\approx-6\uu{eV}$} corresponding to hybridised $sp^3d^2$ covalent bonding (as demonstrated by the overlap of $s$-, $p-$, and $d$-projected DOS in Fig.~\ref{fig:dos}a), (ii) $-6$ to \rev{$\approx-2\uu{eV}$} corresponding predominantly to hybridised N-$p$--TM-$d$ bonding}, and (iii) $\approx-2$ to $0\uu{eV}$ corresponding to metallic bonding formed by TM $d$ electrons. \rev{The term ``covalent'' in the following analysis refers to the regions (i) and (ii) together (i.e., energy between $-10$ and $\approx-2\uu{eV}$).}

To quantify the effect of alloying elements, we introduce ``centre of mass'' (COM) \cite{Rachbauer2011, Rachbauer2011c} of the above mentioned energy bands, calculated as average energy of the band weighted by DOS:
\begin{equation}
  \mathrm{COM}=\frac{\int_{E_{\min}}^{E_{\max}} \rho(E) E\, \mathrm{d}E}{\int_{E_{\min}}^{E_{\max}} \rho(E)\, \mathrm{d}E}
\end{equation}
where $\rho(E)$ is DOS in the band between $E_{\min}$ and $E_{\max}$. \rev{In assessing the alloying effects, the relative decrease of COM in particular energy range contributes towards lower total energy, and consequently leads to chemical strengthening as discussed in Section~\ref{sec:chemical_strength}. Additionally, each energy range (or band) can be characterised by a ``band width'' (BW) defined as:
\begin{equation}
  \text{BW}=2\sqrt{\frac{\int_{E_{\min}}^{E_{\max}} \rho(E) E^2\, \mathrm{d}E}{\int_{E_{\min}}^{E_{\max}} \rho(E)\, \mathrm{d}E}}\ .
\end{equation}
Such quantity reflects the width of the band weighted by the density of states distribution. Smaller value of BW corresponds to better overlap of the hybridised states, and hence to increased degree of hybridisation.} 

COM of the metallic range of Ti$_{0.5}$Al$_{0.5}$N is $-0.684\uu{eV}$. This value changes to $-0.663$, $-0.688$, $-0.768$, $-0.662$, and $-0.762\uu{eV}$ for Y, Zr, Nb, Hf, and Ta, respectively, as evaluated for the Al-excess compositions (see Fig.~\ref{fig:dos}b). Here, a significant strengthening is predicted by alloying of Nb and Ta, related to an increased concentration of $d$ electrons forming the $d$--$d$ bonds ($t_{2g}$ symmetry) along $\langle110\rangle$ directions \cite{Rachbauer2011, Rachbauer2011c}. The same analysis of COM for the covalent region yields $-5.148\uu{eV}$ for Ti$_{0.5}$Al$_{0.5}$N, and $-5.034$, $-5.173$, $-5.192$, $-5.209$, and $-5.256$ for Y, Zr, Nb, Hf, and Ta, respectively, in the Al-excess configurations (Fig.~\ref{fig:dos}b). These numbers suggest that Y considerably weakens the covalent interaction, while the other four elements (Zr, Nb, Hf, and Ta) all strengthen the covalent bonding. \rev{Finally, a consistent picture is obtained also from BW of the hybridised region: Y significantly broadens the electronic states distribution while Ta confines the states resulting in an increased degree of the $sp^3d^2$ hybridisation.}

\rev{In summary,} the most significant strengthening is obtained by alloying with Ta, followed by Nb and Hf. Still a small strengthening is obtained by alloying with Zr, while Y softens the material. These trends are in line with the chemical strength considerations based on the total energy changes, as discussed in Section~\ref{sec:chemical_strength}.

The density of states at the Fermi level is an indicator of the (relative) alloy stability. It reaches a value of $0.21\uu{st.}\cdot\mbox{eV}^{-1}\cdot\mbox{at.}^{-1}$ for Ti$_{0.5}$Al$_{0.5}$N. Zr keeps it unaltered for the Al-excess composition, while Hf, Nb, Ta, and Y increase it to $0.24$, $0.25$, $0.25$, and $0.29\uu{st.}\cdot\mbox{eV}^{-1}\cdot\mbox{at.}^{-1}$, suggesting a gradual destabilisation of Ti$_{1-x}$Al$_x$N by alloying. This is in qualitative agreement with the trends in energy of formation as shown in Fig.~\ref{fig:Ef}.

\subsubsection{Decomposition of the unstable alloys}\label{sec:decomposition}

The mixing enthalpy, $H_{\mathrm{mix}}$, is calculated as
\begin{multline}
  H_{\mathrm{mix}}=E(\mbox{\TiAlXN{X}})-\Big[(1-x-y)E(\mathrm{TiN})\\ +xE(\mathrm{AlN})+yE(\mathrm{XN})\Big]\ ,
\end{multline}
where $E(XN)$ is the total energy (or energy of formation, or cohesive energy) of binary cubic XN, and $E(\mbox{\TiAlXN{X}})$ is the corresponding energy of the quaternary alloy. The mixing enthalpy expresses the energy gain (when negative) of forming the alloy with respect to the binary nitrides, hence it quantifies whether the alloy is stable ($H_{\mathrm{mix}}<0$) or unstable ($H_{\mathrm{mix}}>0$) with respect to decomposition into the cubic binary nitrides.

\begin{figure*}
  \centering
  \includegraphics{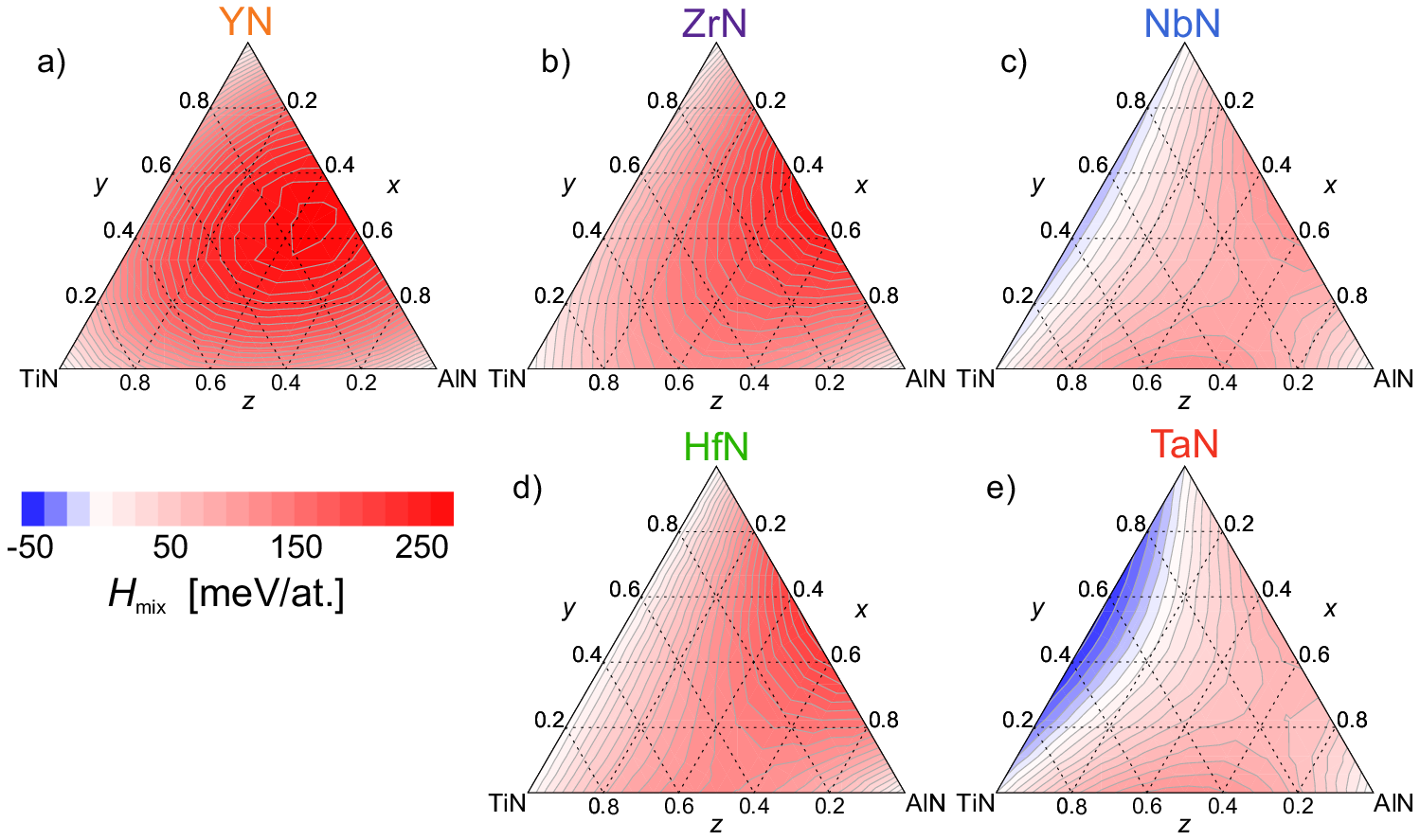}
  \caption{Mixing enthalpy, $H_{\mathrm{mix}}$, as a function of the composition of pseudo-quaternary alloys: (a) \TiAlXN{Y}, (b) \TiAlXN{Zr}, (c) \TiAlXN{Nb}, (d) \TiAlXN{Nb}, and (e) \TiAlXN{Ta}. Contours are every $50\uu{meV/at.}$.}\label{fig:Hmix}
\end{figure*}

The results obtained are plotted in Fig.~\ref{fig:Hmix} for the whole compositional range $0\leq x\leq1$, $0\leq y\leq1$, and $0\leq z\leq1$, $x+y+z=1$, for all five quasi-ternary systems. The driving force for decomposition is the largest for \TiAlXN{Y} near to the YN--AlN tie-line (Fig.~\ref{fig:Hmix}a). The lattice difference between YN and AlN is about $19\%$, a value larger than what is ``allowed'' for solid solutions by Hume-Rothery rules \cite{Cottrell1967}. $H_{\mathrm{mix}}$ is positive in the whole compositional range except for alloys close to the quasi-binary TiN--NbN and TiN--TaN tie-line (Figs.~\ref{fig:Hmix}c and e). This suggests that the Ti$_{1-y}$Nb$_y$N and Ti$_{1-y}$Ta$_y$N alloys are stable, a result previously shown in literature \cite{Matenoglou2009, Rachbauer2010, Sangiovanni2011}. The calculated mixing enthalpies are summarised in Table~\ref{tab:data}.

Adding a small amount (up to $y=0.1$) of Y, Zr, or Hf to Ti$_{0.5}$Al$_{0.5}$N while either keeping the Al-to-Ti ratio, or keeping the Al amount $x=0.5$ constant, the driving force for decomposition $H_{\mathrm{mix}}$ increases. Ti$_{1-x}$Al$_x$N has been shown to decompose spinodally before the stable wurtzite AlN precipitates appear \cite{Rachbauer2011a}, and also the quaternary alloys are expected to do so. Therefore, the increased force for decomposition is interpreted as increased force for the isostructural decomposition, thus shifting the onset of spinodal decomposition to lower temperatures (lower thermal loads). Consequently, age-hardening of these alloys is predicted to occur at lower temperatures than for Ti$_{0.5}$Al$_{0.5}$N. The peak broadening in X-ray diffractograms suggesting the spinodal decomposition is indeed observed for lower annealing temperatures when Zr \cite{Chen2011} is added . On the contrary, addition of Nb and Ta lowers the mixing enthalpy thus the onset of the spinodal decomposition is predicted to be shifted to slightly higher thermal loads as compared with Ti$_{0.5}$Al$_{0.5}$N, as in excellent agreement with experiments, see e.g. Ref.~\cite{Rachbauer2011c}.

\section{Conclusions}

We employed first principle calculations to study alloying effects of early transition metals on the ground state properties and stability of Ti$_{1-x}$Al$_x$N alloys used as a protective hard coating material. The calculated lattice parameters of the quaternary alloys exhibit a deviation from Vegard's-like linear relationship, being the largest with $\approx2.5\%$ for Y$_{0.5}$Al$_{0.5}$N. The additional compositional degree of freedom of the quaternary alloys (as compared with Ti$_{1-x}$Al$_x$N) is suggested to allow for specially designed lattice-matched multilayers with alternating soft and hard components.

Analysis of the chemical strength and local elastic distortions showed that Ta and Nb are the most promising strengtheners, closely followed by Zr and Hf. These conclusions agree well with the published experimental results of the increased hardness of Ti$_{1-x}$Al$_x$N alloyed with Zr, Nb, Hf, or Ta. Finally, the analysis of the mixing enthalpy as a measure of the driving force for decomposition suggests that addition of Y, Zr and Hf leads to its increase, thus earlier onset of the spinodal decomposition and the related age-hardening process.

\section*{Acknowledgements}

Financial support by the START Program (Y371) of the Austrian Science Fund (FWF) is greatly acknowledged.

\appendix

\section{Properties of the quasi-ternary alloys}
\begin{sidewaystable*}
%   \scriptsize
  \vspace*{4cm}
  \begin{tabular}{ccc|ccc|ccc|ccc|ccc|ccc}

        &               &               & \multicolumn{3}{c|}{Ti$_z$Al$_x$Y$_y$N}       & \multicolumn{3}{c|}{Ti$_z$Al$_x$Zr$_y$N}      &\multicolumn{3}{c|}{Ti$_z$Al$_x$Nb$_y$N}       & \multicolumn{3}{c|}{Ti$_z$Al$_x$Hf$_y$N}      & \multicolumn{3}{c|}{Ti$_z$Al$_x$Ta$_y$N}       \\
$z$     &      $x$      &       $y$     &       $a$     &       $B$  &$H_\mathrm{mix}$&       $a$     &       $B$  &$H_\mathrm{mix}$&       $a$     &       $B$  &$H_\mathrm{mix}$&       $a$     &       $B$  &$H_\mathrm{mix}$&       $a$     &       $B$  &$H_\mathrm{mix}$ \\
        &               &               &     [\AA]     &      [GPa]    &     [meV/at.] &     [\AA]     &      [GPa]    &     [meV/at.] &     [\AA]     &      [GPa]    &     [meV/at.] &     [\AA]     &      [GPa]    &     [meV/at.] &     [\AA]     &      [GPa]    &     [meV/at.] \\\hline
0.0     &       0.0     &       1.0     &       4.918   &       159     &       0       &       4.621   &       245     &       0       &       4.458   &       305     &       0       &       4.538   &       269     &       0       &       4.421   &       336     &       0       \\
0.0     &       0.2     &       0.8     &       4.828   &       135     &       176     &       4.545   &       236     &       160     &       4.398   &       268     &       74      &       4.473   &       252     &       136     &       4.378   &       279     &       54      \\
0.0     &       0.4     &       0.6     &       4.670   &       158     &       264     &       4.453   &       233     &       240     &       4.339   &       247     &       110     &       4.394   &       241     &       203     &       4.312   &       276     &       82      \\
0.0     &       0.6     &       0.4     &       4.527   &       159     &       264     &       4.350   &       233     &       240     &       4.265   &       253     &       112     &       4.308   &       240     &       203     &       4.251   &       259     &       82      \\
0.0     &       0.8     &       0.2     &       4.297   &       216     &       176     &       4.228   &       237     &       160     &       4.185   &       248     &       75      &       4.205   &       241     &       136     &       4.177   &       240     &       54      \\
0.0     &       1.0     &       0.0     &       4.070   &       253     &       0       &       4.070   &       253     &       0       &       4.070   &       253     &       0       &       4.070   &       253     &       0       &       4.070   &       253     &       0       \\
0.2     &       0.0     &       0.8     &       4.806   &       140     &       94      &       4.560   &       242     &       35      &       4.425   &       296     &       -12     &       4.495   &       258     &       2       &       4.399   &       315     &       -33     \\
0.2     &       0.2     &       0.6     &       4.683   &       60      &       236     &       4.470   &       236     &       151     &       4.357   &       279     &       49      &       4.419   &       253     &       104     &       4.336   &       300     &       20      \\
0.2     &       0.4     &       0.4     &       4.564   &       60      &       278     &       4.373   &       233     &       197     &       4.290   &       263     &       83      &       4.336   &       245     &       150     &       4.275   &       277     &       55      \\
0.2     &       0.6     &       0.2     &       4.363   &       94      &       229     &       4.263   &       232     &       173     &       4.217   &       254     &       92      &       4.242   &       243     &       141     &       4.208   &       262     &       73      \\
0.2     &       0.8     &       0.0     &       4.124   &       252     &       80      &       4.124   &       252     &       78      &       4.124   &       252     &       75      &       4.124   &       252     &       77      &       4.124   &       252     &       74      \\
0.4     &       0.0     &       0.6     &       4.673   &       155     &       141     &       4.493   &       245     &       52      &       4.388   &       279     &       -17     &       4.445   &       260     &       3       &       4.371   &       269     &       -49     \\
0.4     &       0.2     &       0.4     &       4.547   &       60      &       241     &       4.398   &       244     &       135     &       4.318   &       276     &       40      &       4.365   &       254     &       83      &       4.303   &       291     &       11      \\
0.4     &       0.4     &       0.2     &       4.348   &       95      &       233     &       4.293   &       245     &       156     &       4.248   &       264     &       83      &       4.274   &       253     &       121     &       4.240   &       272     &       64      \\
0.4     &       0.6     &       0.0     &       4.162   &       260     &       121     &       4.162   &       260     &       118     &       4.162   &       260     &       113     &       4.162   &       260     &       116     &       4.162   &       260     &       111     \\
0.6     &       0.0     &       0.4     &       4.546   &       160     &       141     &       4.421   &       256     &       52      &       4.346   &       276     &       -17     &       4.390   &       264     &       3       &       4.335   &       269     &       -49     \\
0.6     &       0.2     &       0.2     &       4.378   &       95      &       188     &       4.319   &       259     &       110     &       4.275   &       280     &       49      &       4.301   &       265     &       74      &       4.268   &       286     &       29      \\
0.6     &       0.4     &       0.0     &       4.196   &       269     &       121     &       4.196   &       269     &       118     &       4.196   &       269     &       113     &       4.196   &       269     &       116     &       4.196   &       269     &       111     \\
0.8     &       0.0     &       0.2     &       4.401   &       219     &       94      &       4.342   &       271     &       35      &       4.302   &       291     &       -12     &       4.327   &       274     &       2       &       4.296   &       295     &       -33     \\
0.8     &       0.2     &       0.0     &       4.227   &       281     &       80      &       4.227   &       281     &       78      &       4.227   &       281     &       75      &       4.227   &       281     &       77      &       4.227   &       281     &       74      \\
1.0     &       0.0     &       0.0     &       4.256   &       292     &       0       &       4.256   &       292     &       0       &       4.256   &       292     &       0       &       4.256   &       292     &       0       &       4.256   &       292     &       0       \\
  \end{tabular}
  \caption{Calculated lattice parameter, $a$, bulk modulus, $B$, and mixing enthalpy, $H_{\mathrm{mix}}$, of quasi-ternary TiN--AlN--XN systems for $X=$Y, Zr, Nb, Hf, and Ta, as a function of TiN ($z$), AlN ($x$) and XN ($y$) mole fractions.}\label{tab:data}
\end{sidewaystable*}

% \bibliographystyle{aipnum4-1}
% \bibliography{quaternaries}


\end{document}